\documentclass[12pt]{iopart}
\usepackage{iopams}

\usepackage{graphicx}
\usepackage{amsthm}
\usepackage{amstext}
\newcommand{\RM}{\mathbb{R}}

\newcommand{\NM}{\mathbb{N}}

\newtheorem{theorem}{Theorem}
\newtheorem{lemma}{Lemma}
\newtheorem{corollary}{Corollary}
\newtheorem{proposition}{Proposition}

\begin{document}

\title[Classical limit of the quantum Zeno effect]{Classical limit of the quantum Zeno effect}

\author{Paolo Facchi$^{1,2}$,  Sandro Graffi$^{3}$, Marilena Ligab\`o$^{1}$,}

\address{$^1$Dipartimento di Matematica, Universit\`a di Bari, I-70125 Bari, Italy}
\address{$^2$Istituto Nazionale di Fisica Nucleare, Sezione di Bari, I-70126 Bari, Italy}
\address{$^3$Dipartimento di Matematica, Universit\`{a} di Bologna, I-40127 Bologna, Italy}

\begin{abstract}
The evolution of a quantum system subjected to infinitely many measurements in a finite time interval is confined in a proper subspace of the Hilbert space. This phenomenon is called ``quantum Zeno effect": a particle under intensive observation does not evolve. This effect is at variance with the classical evolution, which obviously is not affected by any observations. By a semiclassical analysis we will show that the quantum Zeno effect vanishes at all orders, when the Planck constant tends to zero, and thus it is a purely quantum phenomenon without classical analog, at the same level of tunneling.
\end{abstract}

\date{\today}
\maketitle

%%%%%%%%%%%%%%%%%%%%%%%%%%%%%%%%%%%%%%%

\maketitle

\section{Introduction}
In this paper we study the classical limit of the quantum Zeno
effect in its simplest formulation, namely a free particle subjected to position measurements. The presence of any smooth and bounded potential does not affect our results.

Therefore, let us consider a free quantum particle in $\RM^n$. Its states are
described by vectors in the Hilbert space $\mathcal{H}=L^2(\RM^n)$ and the Schr\"odinger operator is $H=-\hbar^2 \Delta/2m$ with domain the Sobolev space $D(H)=H^2(\RM^n)$.
Let $P=\chi_{\Omega}$ be the orthogonal projection onto a compact set $\Omega \subset \RM^n$ with regular boundary. Here $\chi_{\Omega}$ denotes the characteristic function of the set $\Omega$ ($\chi_{\Omega}(x)$ equals $1$ for $x\in\Omega$ and $0$ otherwise). $P$ is the observable associated to a measurement that ascertains whether or not the particle is in the spatial region $\Omega$.
If one performs $N$ measurements on the particle at regular time intervals of length $t/N$, at the end of this procedure the state of the system is, up to a normalization,
\begin{equation}
\psi_N(t)=(P\, U(t/N)\, P)^N\psi,
\end{equation}
where $\psi$ is the initial state of the particle and $U(t)=e^{-itH/\hbar}$ is the evolution group generated by $H$.
Let
\begin{equation}
V_{N}(t)=(Pe^{-i t H /\hbar N}P)^N.
\label{eq:productformula}
\end{equation}
We are interested in the limit $N \to +\infty$ of the product formula $V_{N}(t)$. In Ref.~\cite{exner} it has been proved that
\begin{theorem}
There exists a set $M \subset \RM$ of Lebesgue measure zero and a strictly increasing sequence $\{N\}$ of positive integers along which we have
\label{lim. Zeno}
\begin{equation*}
\lim_{N \to +\infty}V_{N}(t) \psi =e^{-itH_{\Omega}/\hbar} P \psi,
\end{equation*}
for all $\psi \in \mathcal{H}$ and  for all $t \in \RM \setminus M$, where $H_{\Omega}=-\hbar^2 \Delta_{\Omega}/2m$, and $\Delta_{\Omega}$ is the Laplace operator with Dirichlet boundary condition on $\partial \Omega$, that is $D(H_{\Omega}) = H^2(\Omega)\cap H^1_0 (\Omega)$.
\end{theorem}
The limit in Theorem~\ref{lim. Zeno} implies that, if it is possible to perform infinitely many position measurements in the finite time interval $[0,t]$ the probability of finding the particle in the region $\Omega$ in \emph{each} of these measurements reads
\begin{equation}
p_N(t) = \langle \psi_N(t),  \psi_N(t)\rangle = \| V_N(t) \psi \|^2 \to \| P \psi \|^2 =1,
\label{eq:QZE}
\end{equation}
for $N\to+\infty$.
This peculiar quantum behavior was named \emph{quantum Zeno effect} by Misra and Sudarshan \cite{misra}.
Since then,  the quantum Zeno effect has received constant attention by physicists and mathematicians. For an up-to-date review of the main mathematical and physical aspects, see \cite{ZenoMP} and references therein.

The effect has been observed  experimentally in a variety of systems, on
experiments involving photons \cite{kwiat}, nuclear spins \cite{Chapovsky}, ions \cite{balzer2002},  optical pumping \cite{molhave2000}, photons in a cavity \cite{haroche}, ultracold atoms \cite{raizenlatest} and Bose-Einstein condensates \cite{ketterle}. Moreover, these ideas might lead to remarkable applications, e.g.\ in quantum computation and in the control of decoherence.

Of course, the behavior in (\ref{eq:QZE}) is at complete variance with that of a classical particle. Indeed, a free particle with a nonzero initial momentum will eventually escape from the region
$\Omega$ and obviously  its motion is not modified by any observations.
More precisely, a particle with initial momentum $\xi\neq 0$ after a time
\begin{equation}
T_\xi = m \delta(\Omega)/ |\xi|
\end{equation}
will be surely found outside $\Omega$, independently of its initial position $x\in\Omega$, where
\begin{equation}
\delta(\Omega)=\sup_{x,y\in\Omega} |x-y|
\end{equation}
is the diameter of $\Omega$.

In this paper we will prove that the quantum Zeno effect is a purely quantum phenomenon, at the same level of tunneling; namely, it  cannot be observed at any finite order in $\hbar$, in the limit $\hbar\to 0$.
Notice that, in order to compare classical and quantum dynamics one has to describe them in the same space. In fact, by the Wigner--Moyal formalism, one can give a description of quantum mechanics in classical phase space, which is completely equivalent  to the usual description in Hilbert space. Functions $\tau(x,\xi)$ on the phase space $\mathbb{R}^{2n}$ (classical observables) are mapped into operators $T=\textrm{Op}^{W}(\tau)$ on the Hilbert space  $L^2(\mathbb{R}^{n})$ (quantum observables) via the Weyl quantization map.
In particular,  the noncommutative product of two quantum operators $T_1 T_2$ corresponds to the twisted convolution product  (definition recalled below) $\tau_1\sharp\tau_2$ of the classical observables $\tau_1$ and $\tau_2$, while the commutator $[T_1,T_2]$ corresponds  to the Moyal bracket $\{\tau_1,\tau_2\}_M$.   The main point here is that both  $\tau_1\sharp\tau_2$ and $\{\tau_1,\tau_2\}_M$ depend on the Planck constant  $\hbar$. When $\hbar\to 0$ they reduce to commutative multiplication and Poisson bracket, respectively, thus restoring  classical mechanics. Semiclassical analysis deals with all quantum corrections to classical mechanics at each order in $\hbar$, which are encoded in asymptotic power series in $\hbar$ of $\tau_1\sharp\tau_2$ and $\{\tau_1,\tau_2\}_M$. 

In the following we will analyze (a suitable regularization of) the product formula  in (\ref{eq:productformula}) with the above-mentioned tools.
Let
\begin{equation}
\tilde{V}_N(t)= P_N(t)\, P_N\left( \frac{N-1}{N}t\right) \dots P_N\left( \frac{2}{N}t\right) P_N\left( \frac{1}{N}t\right) P_N(0),
\label{eq:regproductformula}
\end{equation}
where $P_{N}$ is the multiplication operator by a suitable $C^{\infty}$ mollification of the characteristic function $\chi_{\Omega}$ (see section \ref{sec:reg. proj.}) and $P_{N}(s)$, $s \in \RM$, is the evolution of $P_{N}$ in the Heisenberg picture. Our goal is to prove the following 
\begin{theorem}\label{main th.}
Let $M \subset \RM$, $\{N\}$ and $H_{\Omega}$ be as in the Theorem \ref {lim. Zeno}.
\begin{enumerate}
  \item
       \begin{equation*}
       \lim_{N \to +\infty}\tilde{V}_{N}(t) \psi= e^{itH/\hbar} e^{-itH_{\Omega}/\hbar} P \psi,
       \end{equation*}
       for all $\psi \in \mathcal{H}$ and for all $t \in \RM \setminus M$.
  \item $\tilde{V}_{N}(t)$ has a semiclassical symbol $\Theta_N $ and
       \begin{eqnarray*}
        \Theta_N (x,\xi;t) \sim \sum_{j=0}^{+\infty} \hbar^j \,\Theta_{j,N} (x,\xi;t) ,
       \end{eqnarray*}
       for $\hbar\to0$ and for  all $x , \xi \in \RM^n$.
  \item For every $\xi \in \RM^{n}$, $\xi\neq0$, if $t > T_{\xi}:= m \delta(\Omega)/ |\xi|$, one has
       \begin{equation*}
       \lim_{N \to +\infty}\Theta_{j,N}(x,\xi;t)= 0,
       \end{equation*}
       uniformly in $x \in \RM$ and $j \in \NM$.
\end{enumerate}
\end{theorem}

Statement (i) of Theorem~\ref{main th.} allows one to replace the product formula (\ref{eq:productformula}) with its regularized version (\ref{eq:regproductformula}), which is more suitable to a semiclassical analysis. Notice, indeed, that, for any $\psi\in\mathcal{H}$, also $\tilde{\psi}_N(t) =\tilde{V}_N(t)\psi$ satisfies Eq.~(\ref{eq:QZE}) and thus is related to the probability of finding the particle in $\Omega$ in all $N$ measurements.

Statement (ii) says that the quantum product formula $\tilde{V}_N(t)$ has a classical counterpart $\Theta_N (x,\xi;t)$ that admits an asymptotic expansion in $\hbar$, and, finally, statement (iii) asserts that each term $\Theta_{j,N}$ of the expansion identically vanishes for $t>T_\xi$ in the limit $N\to\infty$.

This last statement is the main result of this paper. Its physical meaning is the following: we consider the asymptotic expansion of the product formula for $\hbar\to 0$.
About the zero-th order, classical, term we have already discussed: given an initial momentum $\xi\neq 0$, at times $t>T_\xi$ we get $\lim_{N\to \infty} \Theta_{0,N} (x,\xi;t) = 0$, uniformly in $x\in\Omega$, that is the classical particle, initially in $\Omega$,   has eventually escaped from that region. Statement (iii) asserts that the \emph{same} feature is shared by \emph{all} quantum corrections, independently of the order in $\hslash$.

\section{Weyl's quantization and Egorov's theorem}

In this section, mainly intended as a set up of the notation, we will briefly recall the tools needed in the following. For all details and proofs we refer to \cite{Sjostrand, Folland, Robert1, Robert2}.

Let us start with  Weyl's quantization.
Let $\tau$ be a function in the Schwartz space $\mathcal{S}(\RM^{2n})$. We can define the following operator on $L^{2}(\RM^n)$
\begin{equation}
\textrm{Op}^{W}(\tau)=\int_{\RM^{2n}} e^{i(\xi \cdot X+y \cdot p)}\hat{\tau}(\xi,y)\; \frac{d\xi dy} {(2\pi )^n},
\end{equation}
where $X$ is the position operator $X\psi(x)=x\psi(x)$, $p=\hbar D_x/i$ the momentum operator, with $D_x$ the $n$-dimensional gradient, and the Fourier transform is defined by
\begin{equation}
\hat{\tau}(\xi,y)=\int_{\RM^{2n}} \tau(x, \eta) e^{-i(\xi \cdot x+\eta \cdot y)}\; \frac{dx d\eta}{(2\pi)^n}.
\end{equation}
It is easy to check that if $\tau$ is real then $\textrm{Op}^{W}(\tau)$ is a bounded self-adjoint operator.
The operator $\textrm{Op}^{W}(\tau)$ is called the (Weyl) quantization of the symbol $\tau$. Physically, it is interpreted as the quantum observable corresponding to the classical observable $\tau$.

One can prove that, for any $\psi\in\mathcal{S}(\RM^{n})$
\begin{equation}
(\textrm{Op}^{W}(\tau)\psi)(x)=\int_{\RM^{2n}}
  \tau\left(\frac{x+y}{2},\xi\right)e^{-i\xi\cdot(y-x)/\hbar}\psi(y) \;\frac{d\xi dy}{(2\pi \hbar)^n}.
\label{eq:intkern}
\end{equation}
Equation (\ref{eq:intkern}) allows one to extend the quantization map to tempered distributions $\tau$.
We also recall the definition of the twisted convolution product between two symbols $\tau_{1}$ and $\tau_{2}$\begin{equation}
\fl \qquad \tau_{1} \sharp \tau_{2} (x,\xi)=\int_{\RM^{4n}}
\tau_{1}(x_1,\xi_1)\tau_{2}(x_2,\xi_2) e^{\frac{2i}{\hbar}[(x-x_1)\cdot(\xi-\xi_2)-(x-x_2)\cdot(\xi-\xi_1)]}\; \frac{dx_1 d\xi_1 dx_2 d\xi_2}{(\pi \hbar)^{2n}}.
\label{eq:twistedconv}
\end{equation}
The twisted product is the image on the space of symbols of the noncommutative operator product, namely
\begin{equation}
\textrm{Op}^{W}(\tau_{1}\sharp \tau_{2})=\textrm{Op}^{W}(\tau_{1})\, \textrm{Op}^{W}(\tau_{2}).
\end{equation}

The last ingredient we need in our analysis is Egorov's theorem that tell us how the time evolution and the Weyl quantization are related.
We will focus our attention to the case we are interested in, i.e.\  the free Hamiltonian.
In this case the Schr\"odinger operator $H=\textrm{Op}^{W}(\mathcal{H})$ is the Weyl quantization of the Hamiltonian $\mathcal{H}(x,\xi)=\xi^2/2m$.
The time evolution of a classical bounded observable is
$\tau_{t}:=\tau \circ \phi^{\mathcal{H}}_{t}$, where
\begin{equation}
\phi^{\mathcal{H}}_{t}:\RM^{2n}\to \RM^{2n} \quad \phi^{\mathcal{H}}_{t}(x,\xi)=\left(x+\frac{\xi t}{m}, \xi\right)
\end{equation}
is the Hamiltonian flow.
On the other hand, the quantum time evolution of a bounded observable $T=\textrm{Op}^{W}(\tau)$ is
\begin{equation}
T(t)=e^{itH/\hbar}Te^{-itH/\hbar},
\end{equation}
which is a solution of the equation
\begin{equation}\label{eq.quant.}
\dot{T}(t)=\frac{i}{\hbar}[H,T(t)] .
\end{equation}
Let $\tau_{t}^{\hbar}(x,\xi)$ be the symbol of $T(t)$, namely $T(t)=\textrm{Op}^{W}(\tau_{t}^{\hbar})$.
Equation (\ref{eq.quant.}) is mirrored into the following equation for the symbol $\tau_{t}^{\hbar}$ on the phase space
\begin{equation}\label{eq.cl.}
\dot{\tau}_{t}^{\hbar}=\{\mathcal{H}, \tau_{t}^{\hbar}\}_{M}
\end{equation}
with the initial condition $\tau_{0}^{\hbar}= \tau $, where
\begin{equation}
\{f,g\}_{M}=f \sharp g- g\sharp f ,
\end{equation}
is the Moyal bracket.
Solving equation (\ref{eq.cl.}) one finds that, since $\mathcal{H}$ is quadratic in $(x,\xi)$
\begin{equation}\label{eg.th.}
\tau_{t}^{\hbar}= \tau_{t}= \tau\circ \phi^{\mathcal{H}}_{t},
\end{equation}
namely
\begin{equation}
T(t)=e^{itH/\hbar}Te^{-itH/\hbar}=\textrm{Op}^{W}(\tau_{t}^{\hbar})=\textrm{Op}^{W}(\tau \circ \phi^{\mathcal{H}}_{t}).
\end{equation}
Thus, in this case time evolution and quantization commute. For general non quadratic Hamiltonian, the semiclassical Egorov theorem (see  \cite{Robert1}) states that (\ref{eg.th.}) holds only at order $0$ in $\hbar$.

\section{A modified product formula}\label{sec:reg. proj.}

\begin{figure}
\begin{center}
\includegraphics[width=0.7\textwidth]{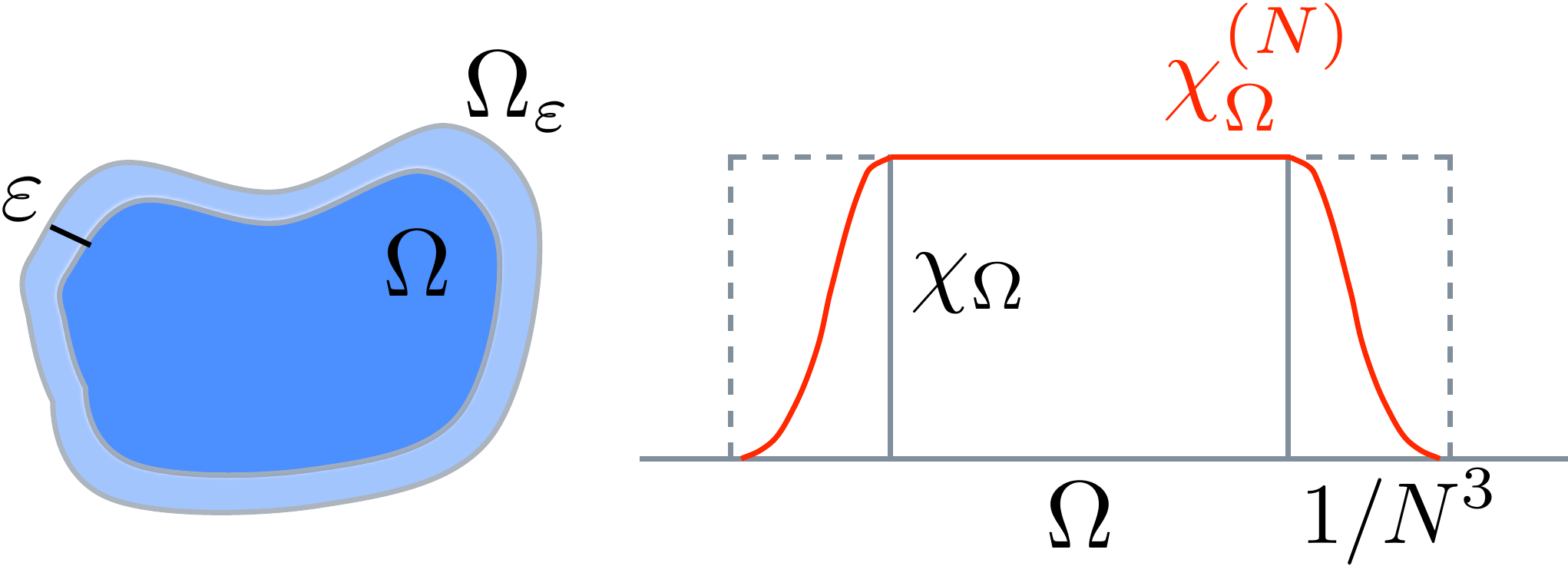}
\caption{$\varepsilon$-neighborhood of the compact set $\Omega$ and mollified characteristic function $\chi_{\Omega}^{(N)}$, with $\varepsilon_N=1/N^3$.}
\label{fig:mollified}
\end{center}
\end{figure}

The projection $P$ can be considered as a pseudodifferential operator whose symbol is the characteristic function $\varsigma(x,\xi)=\chi_{\Omega}(x)$ of the set $\Omega\times\RM^n$ in the phase space.
However, in order to have a sufficiently smooth symbol, instead of the projection $P$, we consider an operator $0\leq P_{N}\leq 1$ as the Weyl quantization of a symbol  which is a $C^{\infty}$ mollification of the characteristic function $\chi_{\Omega}$. Namely, given an $\varepsilon$-neighbourhood of the domain $\Omega$
\begin{equation}
\Omega_{\varepsilon}=\{x\in\RM^n \,|\, d(x,\Omega)<\varepsilon\},
\end{equation}
with $d(x,\Omega)=\inf_{y\in\Omega} |x-y|$ (see Fig.~\ref{fig:mollified}),
we take
\begin{equation}
P_{N} = \textrm{Op}^{W}(\vartheta^{(N)}), \qquad \vartheta^{(N)} (x,\xi)= \chi^{(N)}_\Omega(x) ,
\label{eq:tildeP}
\end{equation}
where
\begin{equation}
 \chi_\Omega \leq \chi^{(N)}_\Omega \leq  \chi_{\Omega_{1/N^3}}, \qquad
 \chi^{(N)}_\Omega \in C^{\infty}(\RM^{n})
\label{eq:chiN}
\end{equation}
is a smoothed approximation of the characteristic function $\chi_\Omega$ supported in $\Omega_{\varepsilon_N}$, with $\varepsilon_N=1/N^3$. See Fig.~\ref{fig:mollified}.

Observe now that,  since $N(P_{N}-P) u \to 0$  when $N\to\infty$, one has that 
\begin{equation}
N(P_N e^{-it H/\hbar N}P_N - P e^{-it H/\hbar N}P) u \to 0,
\end{equation}
for any $u$ such that $u, Pu \in H^2(\RM^n)$, which is a dense subset of $L^2(\RM^n)$. Therefore, the limit generators of the two discrete semigroups coincide.
By Theorem~\ref{lim. Zeno} it follows that 
\begin{lemma}
Let $M$, $\{N\}$ and $H_{\Omega}$ be as in the Theorem \ref {lim. Zeno}. One has
\label{lemma1}
\begin{equation*}
\lim_{N\to +\infty} \left(P_{N}e^{-itH/\hbar N}P_{N}\right)^{N} \psi
=e^{-itH_{\Omega}/\hbar} P\psi,
\end{equation*}
for all $\psi \in \mathcal{H}$ and for all $t \in \RM \setminus M$.
\end{lemma}
\noindent Therefore, in our analysis of the quantum Zeno effect we can use the sequence $\{P_{N}\}$ in place of the projection $P$. 
Note that, while the projection $P$ is associated with a yes/no spatial measurement  which ascertains whether or not the particle is  in the region $\Omega$, its smoothed version $P_{N}$ corresponds to a fuzzy spatial measurement which is not sharp at the boundary of the region.  Thus, the  physical meaning of the above statement is that the interference effects arising from a small smoothing of the projection do not affect the overall phenomenon.

First let us rewrite $V_{N}(t)$ in a more convenient way.
By using the evolution of $P$ in the Heisenberg picture,
\begin{equation}
P(s)=e^{isH/\hbar}Pe^{-isH/\hbar},
\end{equation}
we obtain
\begin{eqnarray}
V_{N}(t) & = & (Pe^{-itH/N\hbar}P)^N \nonumber\\
         & = & e^{-itH/\hbar}P(t)\,P\left( \frac{N-1}{N}t\right) \dots P\left( \frac{2}{N}t\right) P\left( \frac{1}{N}t\right) P(0).
\label{eq:VNnew}
\end{eqnarray}
Now let us substitute in the above equation the projection $P$ with the positive operator $P_{N}$ given by Eq.~(\ref{eq:tildeP}) and neglect the final (trivial) unitary evolution in (\ref{eq:VNnew}).  We end up with the following product formula
\begin{equation}\label{op}
\tilde{V}_N(t)= P_N(t)\, P_N\left( \frac{N-1}{N}t\right) \dots P_N\left( \frac{2}{N}t\right) P_N\left( \frac{1}{N}t\right) P_N(0).
\end{equation}
This is the main object of our investigation.

Notice now that from Theorem~\ref{lim. Zeno} and Lemma~\ref{lemma1} we immediately get the following
\begin{corollary}\label{mod.prod.form.}
One gets
\begin{equation*}
\lim_{N \to +\infty}\tilde{V}_{N}(t) \psi= e^{itH/\hbar} e^{-itH_{\Omega}/\hbar} P \psi,
\end{equation*}
for all $\psi \in \mathcal{H}$ and for all $t \in \RM \setminus M$.
\end{corollary}
\noindent This is a reformulation of the quantum Zeno effect: the strong limit of the product formula (\ref{op}) exists and yields a nontrivial evolution. In particular, notice that, for any $\psi\in\mathcal{H}$, also $\tilde{\psi}_N(t) =\tilde{V}_N(t)\psi$ satisfies Eq.~(\ref{eq:QZE}). Observe that Corollary~\ref{mod.prod.form.} is statement $(i)$ of Theorem~\ref{main th.}.

\section{Semiclassical analysis of the quantum Zeno effect}

Now we have all the ingredients to prove the last statements of Theorem~\ref{main th.}. Let us focus  on the classical limit of the product formula (\ref{op}).
First we can construct $\vartheta^{(N)}(x,\xi;t)=(\vartheta^{(N)} \circ \phi^{\mathcal{H}}_{t})(x,\xi)$, which, since the Hamiltonian is quadratic, coincides with the symbol of the Heisenberg evolution of $P_N$.
Define for all $k=0, \dots, N$,
\begin{equation}
\vartheta_{k}(x,\xi):=\vartheta^{(N)}\left(x,\xi; {\frac{k t}{N}}\right), 
\label{eq:thetak}
\end{equation}
so that the symbol of the operator (\ref{op}), $\tilde{V}_N(t)= \textrm{Op}^{W}(\Theta_N)$, is given by
\begin{equation}\label{twisted N convolution}
\Theta_N = \vartheta_{N}\sharp \vartheta_{N-1}\sharp \dots \sharp \vartheta_{1}\sharp \vartheta_{0}.
\end{equation}
From Eq.~(\ref{eq:twistedconv}), it is not difficult to show that \cite{Robert1}
\begin{eqnarray}
\phi_{1} \sharp \phi_{2} & \sim &  \sum_{j=0}^{+\infty}\left(\frac{i \hbar}{2}\right)^j \frac{1}{j!}
\left(D_{x,\phi_{1}}\cdot D_{\xi,\phi_{2}}-D_{x,\phi_{2}}\cdot D_{\xi,\phi_{1}}\right)^{j}\phi_{1} \phi_{2}\nonumber\\
                         &  =   &  \sum_{j=0}^{+\infty}\left(\frac{i \hbar}{2}\right)^j \frac{1}{j!} \phi_{1} \sharp_{j} \phi_{2} ,
\label{eq:asymconv}
\end{eqnarray}
where
\begin{equation}
\phi_{1} \sharp_{j} \phi_{2}:=\left(D_{x,\phi_{1}}\cdot D_{\xi,\phi_{2}}-D_{x,\phi_{2}}\cdot D_{\xi,\phi_{1}}\right)^{j}\phi_{1} \phi_{2}.
\end{equation}
Here, the subscripts $\phi_{1}$ and $\phi_2$ indicate that the differentiation is to be applied only to $\phi_{1}$ or $\phi_2$.

By plugging (\ref{eq:asymconv})  into (\ref{twisted N convolution}) we finally obtain the desired asymptotic power series in $\hbar$ of  the symbol $\Theta_N(t)$ of the product formula $\tilde{V}_N(t)$ in (\ref{op}):
\begin{eqnarray}
\Theta_N (x,\xi;t) \sim \sum_{j=0}^{+\infty}
\hbar^j \,
\Theta_{j,N} (x,\xi;t) ,
\end{eqnarray}
where
\begin{equation}
\Theta_{j,N}:=\frac{i^j}{2^j j!} \sum_{ j_{1},\dots, j_{N}} \delta_{j_{1}+\dots +j_{N},j} \;
\vartheta_{N}\sharp_{j_N}(\vartheta_{N-1}\sharp_{j_{N-1}}( \dots ( \vartheta_{1}\sharp_{j_1} \vartheta_{0}) \dots )) ,
\label{eq:ThetajN}
\end{equation}
with $\delta_{k,l}$ the Kronecker delta.
This is statement $(ii)$ of Theorem~\ref{main th.}.

Observe that $\Theta_{j,N}$ is a function of $x$ and $\xi$ and $t$, namely is a function of the initial position and
momentum of the particle and of the total time of the experiment.
We want to prove that at each order $j$, whatever the initial nonzero momentum, after a certain time the particle is no longer confined in the region of observation. Precisely, we will prove the last statement of Theorem~\ref{main th.}:
\begin{proposition}
For every $\xi \in \RM^{n}$, $\xi\neq0$, 
one gets that, for all $t > T_{\xi}:=m \delta(\Omega)/ |\xi|$,
\begin{equation*}
\lim_{N \to +\infty} \Theta_{j,N}(x,\xi;t)=0,
\end{equation*}
uniformly in $x \in \RM$ and $j \in \NM$.
\end{proposition}
\begin{figure}
\begin{center}
\includegraphics[width=0.48\textwidth]{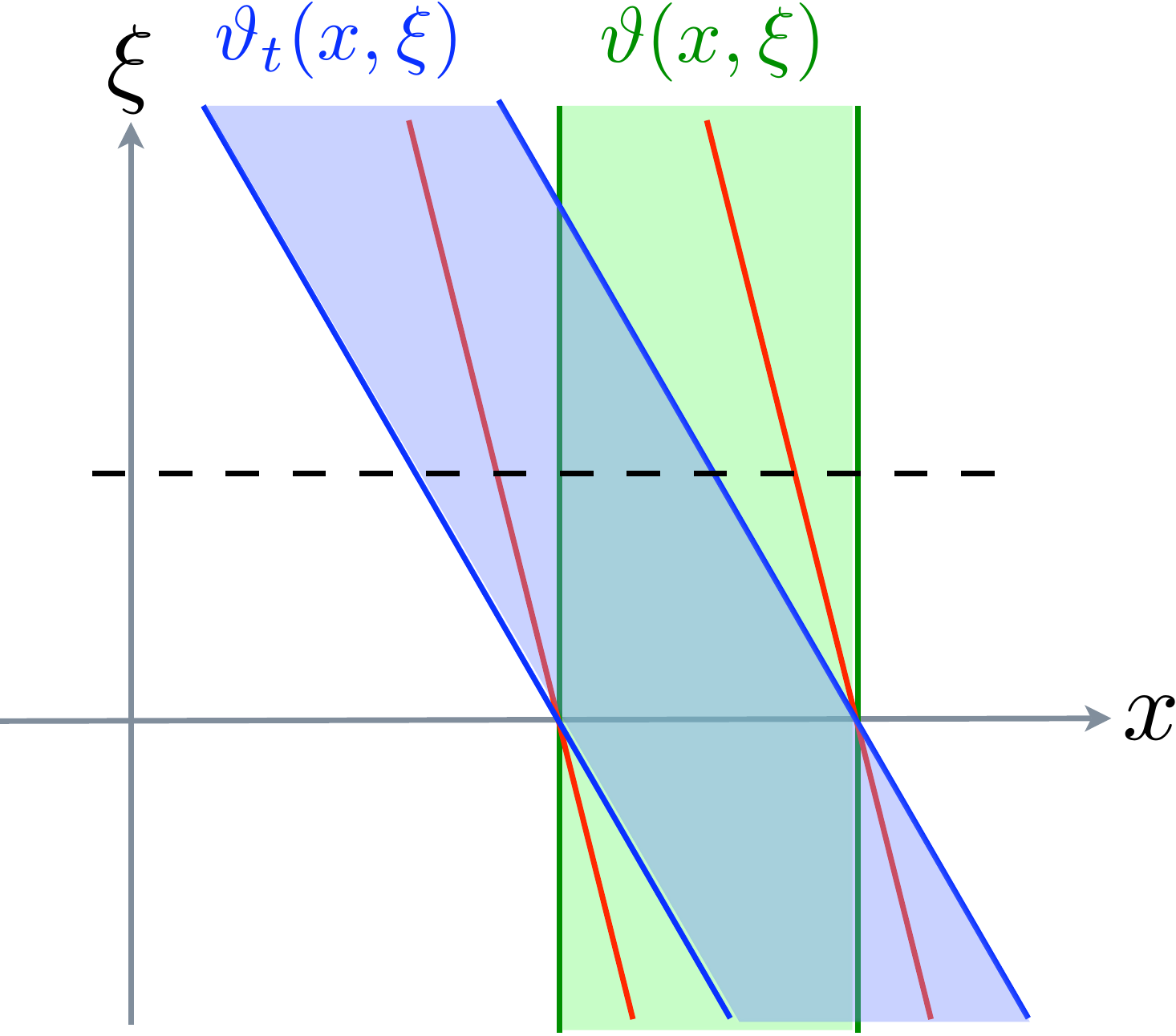}
\caption{Phase space representation of $\Theta_{0,N}(x,\xi)$. The dashed line denotes the initial momentum $\xi\neq0$ of the particle.}
\label{fig:phase_space}
\end{center}
\end{figure}
\begin{proof}
Let us fix the initial momentum of the particle $\xi \neq 0$. 

Consider first the case $j=0$,
\begin{equation}
\Theta_{0,N}=\vartheta_{N}\dots\vartheta_{1}\vartheta_{0}. 
\end{equation}
By making use of (\ref{eq:thetak}) and (\ref{eq:tildeP}) we get
\begin{equation}
\vartheta_{k}(x,\xi) = \chi^{(N)}_\Omega \left(x+ \frac{k t \xi}{Nm} \right),
\end{equation}
so that
\begin{equation}
\Theta_{0,N}(x,\xi;t) = \chi^{(N)}_\Omega (x+\xi t/m)\dots \chi^{(N)}_\Omega (x+\xi t/(mN)) \chi^{(N)}_\Omega (x).
\end{equation}
Since by Eq.~(\ref{eq:chiN}) $\chi^{(N)}_\Omega$ is a mollification of the characteristic function $\chi_\Omega$, we get that the supports satisfy the equation
\begin{equation}
\fl \qquad \mathrm{supp} [\Theta_{0,N}(\cdot,\xi;t)]\subset
\mathrm{supp} [\vartheta_N(\cdot,\xi)\, \vartheta(\cdot,\xi) ]=
\mathrm{supp} [\chi^{(N)}_\Omega (\cdot+\xi t/m)\,\chi^{(N)}_\Omega].
\end{equation}
Observe that
\begin{equation}
\textrm{supp}[\chi^{(N)}_\Omega (\cdot+\xi t/m)]=\textrm{supp}[\chi^{(N)}_\Omega]-\xi t/m \subset \Omega_{\varepsilon_N} -\xi t/m
\end{equation}
by Eq.~(\ref{eq:chiN}).
Therefore, the support
\begin{equation}
\mathrm{supp} [\Theta_{0,N}(\cdot,\xi;t)]\subset \left(\Omega_{\varepsilon_N}-\xi t/m \right)
\end{equation}
is empty if $t> T_{\xi}^N:=m \delta(\Omega_{\varepsilon_N})/ |\xi|$. See Fig.~\ref{fig:phase_space}.
Therefore, since $T_{\xi}^N \to T_{\xi}:=m \delta(\Omega)/ |\xi|$,  we have proved that for any $t > T_{\xi}$ 
\begin{equation}
\Theta_{0,N}(\cdot,\xi;t)\equiv 0,
\end{equation}
 for sufficiently large $N$.

Let us consider now $j > 0$.
Observe that
\begin{equation}
\mathrm{supp} [\Theta_{j,N}(\cdot,\xi;t)]\subset \mathrm{supp} [\Theta_{0,N}(\cdot,\xi;t)]
\end{equation}
therefore, also in this case we have that for $t > T_{\xi}$
\begin{equation}
\Theta_{j,N}(\cdot,\xi;t)\equiv 0,
\end{equation}
for sufficiently large $N$.
\end{proof}

Notice that this result holds for all $N \in \NM$ and $t \in \RM$, thus in particular it holds if we restrict $N$ and $t$ as in the hypothesis of Theorem~\ref{main th.}.

\section{Concluding remarks}\label{sec:conclusion}
We have shown that the quantum Zeno effect vanishes at all orders in $\hbar$, when $\hbar\to0$,
and thus it is a purely quantum phenomenon without classical analog. Remark that, typically, quantum observables  have instead non-zero asymptotic expansions in $\hbar$: elementary examples are  (see e.g.\cite{LL1965}, \S\S 50,51)  the transition probabilities and the Bohr frequency condition. In the first case the asymptotic expansion yields the quantum corrections  to the classical observable evolved along the classical motion, and in the second case all quantum corrections to the classical frequencies.
The quantum Zeno effect is at variance with the above examples. As such, it represents the  counterpart of quantum tunneling through a confining barrier: in the quantum realm the first yields perfect localization, while the latter yields leakage and also the tunnelling amplitude vanishes to all orders in $\hbar$. And conversely in the classical realm. However, the analogy we have drawn is not yet totally symmetric. Indeed,  quantum tunneling is known to be of order $e^{-1/\hbar}$. In this respect it would be very interesting to know whether the quantum Zeno effect is also exponentially vanishing.

\ack This work is partly supported by the European Community
through the Integrated Project EuroSQIP.

\end{document}